\documentclass[12pt]{article}
\usepackage{amsfonts}
\usepackage{amssymb}
\usepackage{amsmath}
\usepackage{graphicx}

\newcommand{\eq}{\begin{eqnarray}}
\newcommand{\eqx}{\end{eqnarray}}
\newcommand{\ba}{\begin{equation}}
\newcommand{\ea}{\end{equation}}
\newcommand{\f}[2]{\frac{#1}{#2}}

\def\la{\label}
\def\nn{\nonumber \\}     
\def\bi{\bibitem}     
     
\def\d{\partial}     
\title{Unified description of Bjorken and Landau 1+1 hydrodynamics}     
     
\author{A.Bialas$^{a}$, R.A.Janik$^{a}$, and R.Peschanski $^b$\thanks{
e-mails:{\tt bialas@th.if.uj.edu.pl},{\tt ufrjanik@th.if.uj.edu.pl},{\tt 
pesch@spht.saclay.cea.fr}}\\ \\
$^a$ \small M.Smoluchowski Institute of Physics, Jagellonian University,\\
\small Reymonta 4, 30-059 Krakow, Poland.\\
$^b$ \small CEA/DSM/SPhT Saclay
Unit\'{e} de Recherche associ{\'e}e au CNRS\\
\small CEA-Saclay, F-91191 Gif/Yvette Cedex, France.}

\date{}
\begin{document} 
     
\maketitle     
     
\begin{abstract} 
We propose a generalization of the Bjorken in-out Ansatz for fluid trajectories 
which, when applied to
the (1+1) hydrodynamic equations,  generates a 
one-parameter family of 
analytic solutions interpolating between the 
boost-invariant Bjorken picture and the non boost-invariant one by Landau. This 
parameter characterises the proper-time scale when the fluid velocities approach 
the in-out Ansatz. We 
discuss the resulting rapidity distribution of entropy 
for various freeze-out conditions and compare it with the original Bjorken and 
Landau  results. 

\end{abstract}     
     
\section {Introduction}     
     
There is an accumulating evidence that hydrodynamics may be relevant for
the description of the medium created in high-energy heavy ion
collisions 
\cite{revhyd}. Indeed, experimental measurements such as the
elliptic flow 
\cite{flow} 
shows the existence of a collective effect on
the produced particles which can be described in terms of a motion of
the fluid. More precisely, numerical simulations of the hydrodynamic
equations describe quite well the distribution of low-$p_\perp$
particles \cite{revhyd}, with an equation of state close to that of a
``perfect fluid'' with a rather low viscosity. This evidence is of course
indirect, since it relies on assumptions about the initial and final
stages of the evolution of the fluid. Thus some doubts can be cast
either on the full thermalization of the medium, or on the
possibility of accounting for some viscosity of the
fluid \cite{therm}.
Also, hydrodynamics are not expected to work for leading 
particles \emph{i.e.} near the kinematic light-cone.
     
Given these objections, it is important to separate precisely
the
consequences of the hydrodynamic flow from those of the initial and
final conditions. From that point of view, it seems tempting to discuss
a simplified picture which can be qualitatively understood in physical
terms. One such simplification, which we are going to follow in this
paper, is the idea that the evolution of the system before freeze-out 
is dominated by the
longitudinal motion \cite{lan,bj} and thus, in fact, the 
%%%%%%%%%%%%%%%%%%%%%%%%%%%%%%%%%%%%%%%%%%%%%%%%%%%%%%%%%%%%%%%%%%%%%%%%%%
hydrodynamic
%%%%%%%%%%%%%%%%%%%%%%%%%%%%%%%%%%%%%%%%%%%%%%%%%%%%%%%%%%%%%%%%%%%%%%%%%%
 transverse
motion can be
neglected or at least factorized out. Thus we shall
consider the $(1\!+\!1)$ dimensional system.

On the theoretical grounds, there are also quite appealing features for
applying hydrodynamic concepts to high-energy heavy-ion reactions. Such
concepts have been already introduced some time ago and find a plausible 
realization
nowadays. The fact that a rather dense medium is created in the first stage of
the collision allows one to admit that the individual partonic or
hadronic degrees of freedom are not relevant during the early evolution
of the medium and justifies its treatment as a fluid. Moreover, the high
quantum occupation numbers allow one to use a classical picture and to
assume that the ``pieces of fluid'' may follow quasi-classical
trajectories in space-time, expressed as an in-out cascade
\cite{gl} with the straight-line trajectories starting at
the origin, i.e.
\ba
y=\eta \label{bgl}
\ea
where
\ba
y=\frac12 \log\left(\frac{E+p}{E-p}\right) \;;\;\;
\eta=\frac12 \log\left(\frac{t+z}{t-z}\right)
\ea
are respectively the rapidity and ``space-time rapidity'' of the piece of the 
fluid.
     
Note, for further use, that (\ref{bgl}) can be rewritten in the form
\ba
2y=\log u^+-\log u^-= \log z^+- \log z^- \label{uz}
\ea where
$u^\pm=e^{\pm y}$ are the light-cone components of the fluid
(four-)velocity and $z_\pm=t\pm z$ are the light-cone kinematical     
variables.      
     
Taking (\ref{bgl}) as the starting point and using the perfect fluid     
hydrodynamics, Bjorken developped in his seminal paper \cite{bj} 
a suggestive (and
very useful in many applications) physical picture of the 
central     
rapidity region of highly relativistic collisions of heavy ions. In this picture 
the 
condition (\ref{bgl}) leads to a     boost-invariant geometry of the expanding 
fluid 
and thus to the central
plateau in the distribution of particles.     
     
It is now experimentally established \cite{RH}, however, that the     
central rapidity region of heavy ion collisions is better described by a     
Gaussian fit with a width proportional to $Y=\log s$, the total rapidity     
range of the secondaries. This finding has renewing interest     
\cite{renewing} for the pioneering hydrodynamic description by Landau     
\cite{lan} where, indeed, a Gaussian-like     
distribution of the fluid was obtained. For the same reason, new families of 1+1 
relativistic hydrodynamic solutions have been recently proposed 
\cite{csorgo,pratt}.   
     
In the present paper we propose to study a generalization of the formula      
(\ref{bgl}) for the classical trajectory which (as we show in the     
following)     
interpolates naturally between the Landau and Bjorken pictures:     
\ba     
2y=\log u^+-\log u^-= \log f_+(z_+) -\log f_-(z_-) \label{yf}     
\ea     
where $f_{\pm} (z_\pm )$ are a priori arbitrary functions. They have to be     
determined from the hydrodynamic equations.

The hydrodynamic equations are rewritten in terms of light- cone     
variables in the next section. The consequences of 
%%%%%%%%%%%%%%%%%%%%%%%%%%%%%%%%%%%%%%%%%%%%%%%%%%%%%%%%%%%%%%%%%%%%%%%%%%
the quasi-classical Ansatz
%%%%%%%%%%%%%%%%%%%%%%%%%%%%%%%%%%%%%%%%%%%%%%%%%%%%%%%%%%%%%%%%%%%%%%%%%%
(\ref{bgl}) and of the generalized one (\ref{yf}) are     
dicussed in Section 3 where also  the corresponding solutions of the 
hydrodynamical equations     
are derived. Various selections of the freeze-out conditions are     
discussed in Section 4. Our conclusions and comments are     
listed in the last section.

\section{Hydrodynamic equations in light-cone variables}

We consider the perfect fluid for which the energy-momentum tensor is
\begin{equation}     
 T^{\mu\nu}= (\epsilon+p)u^{\mu}u^{\nu} - p \eta^{\mu\nu}  \la{T}
\ea     
where $\epsilon$ is the energy density, $p$ is the pressure and      
$u^{\mu}$ is the 4-velocity. We assume that the energy density and pressure
 are related      
by the equation of state:     
\ba     
\epsilon = gp 
\la{state}     
\ea     
 where $1/\sqrt{g}$ is the sound velocity in the liquid.     
Using     
\ba     
u^\pm\equiv u^0\pm u^1=e^{\pm y}     
\ea
and introducing     
\ba     
z_\pm= t\pm z= z^0\pm z^1 =\tau e^{\pm \eta}\; \rightarrow \;
%%%%%%%%%%%%%%%%%%%%%%%%%%%%%%%%%%%%%%%%%%%%%%%%%%%%%%%%%%%%%%%%%%%%%%%%%%
(\f {\d}{\d z^0}\pm \f {\d}{\d z^1})={\scriptstyle\ \frac12}  \f {\d}{\d z^\pm}   
\equiv{ \scriptstyle\ \frac12} \d_\pm     
%%%%%%%%%%%%%%%%%%%%%%%%%%%%%%%%%%%%%%%%%%%%%%%%%%%%%%%%%%%%%%%%%%%%%%%%%% 
\ea     
where $\tau=\sqrt{z_+z_-}$ is the proper time and $\eta$ is the spatial     
rapidity of the fluid,      
the hydrodynamic equations      
     \ba     
\d_\mu T^{\mu\nu} =0     
\ea     
take the form     
\ba     
\d_\pm T^{01}+\frac12\d_+(T^{11}\pm T^{00})     
-\frac12 \d_-(T^{11}\mp T^{00})=0\ .     
\ea     
Using now (\ref{T}) and the equation of state (\ref{state})     
 we deduce from this     
\eq     
g\d_+[\log p]&=&-\frac{ (1\!+\!g)    ^2}2\d_+y-\frac{g^2\!-\!1   
}2e^{-2y}\d_-y\nn     
g\d_-[\log p]&=&\frac{ (1\!+\!g)    ^2}2\d_-y +\frac{g^2\!-\!1   }2e^{2y}\d_+y \ 
. 
\la{ef}     
\eqx     

These are two equations for two unknowns which describe     
 the state of the liquid: the pressure $p$ and the     
rapidity $y$. They should be expressed in terms of the     
positions $z_+,z_-$ in the liquid. Other thermodynamic quantities can be     
obtained from the equation of state (\ref{state}) and the standard     
thermodynamical identities:     
\ba      
p+\epsilon = Ts\;;\;\; d\epsilon = T ds \la{therm}     
\ea     
where we have assumed for simplicity vanishing chemical potential.     
     
The result is      
\ba     
\epsilon =gp=\epsilon_0 T^{g+1}\;;\;\;s=s_0T^g\;\rightarrow \; s\sim     
\epsilon^{g/(g+1)} . \la{esp}     
\ea     
      Note that (\ref{ef}) implies the consistency condition     
\ba     
\d_+\d_- y=\frac{g^2\!-\!1   }{4 (1\!+\!g)    ^2}\ 
\left\{\d_-\d_-[e^{-2y}]-\d_+\d_+[e^{+2y}]\right\}\ .\la{cs}     
\ea     

\section {Generalized in-out Ansatz}     
     
\subsection {Bjorken's in-out Ansatz and boost-invariance}
     
The simplest possibility to describe the expansion of the fluid     
was suggested by Bjorken \cite{bj} who proposed to use the Ansatz (\ref{bgl}) 
in the hydrodynamical context.
Introducing (\ref{bgl}) into (\ref{ef}) we obtain      
\ba     
g\d_+[\log p]=-\frac{1+g}{2z_+}\;;\;\;     
g\d_-[\log p]=-\frac{g+1}{2z_-}     
\ea     
from which we deduce      
\ba     
p=\epsilon\ g^{-1}= p_0\ (z_+z_-)^{-(g+1)/2g} = p_0\ \tau^{-(g+1)/g}\ ,     
\ea     
where $p_0$ is a constant. Thus the system is     
boost-invariant: the pressure     
does not depend neither on $\eta$ nor on $y$. So are $\epsilon$, $s$ and $T$,    
 given by (\ref{esp}).

 \subsection{Beyond boost invariance}

The data on both nucleon-nucleon and nucleus-nucleus collisions
%%%%%%%%%%%%%%%%%%%%%%%%%%%%%%%%%%%%%%%%%%%%%%%%%%%%%%%%%%%%%%%%%%%%%%%%%%
(see, \emph{e.g.} \cite{RH})   
%%%%%%%%%%%%%%%%%%%%%%%%%%%%%%%%%%%%%%%%%%%%%%%%%%%%%%%%%%%%%%%%%%%%%%%%%%
 show     
that the produced system strongly violates boost invariance (except     
perhaps in a narrow region of small c.m. rapidities). It is thus     
necessary to go beyond (\ref{bgl}).
As already indicated in the Introduction, we propose to study -as a simplest 
generalization of (\ref{bgl})- the 
%%%%%%%%%%%%%%%%%%%%%%%%%%%%%%%%%%%%%%%%%%%%%%%%%%%%%%%%%%%%%%%%%%%%%%%%%%
Ansatz   
%%%%%%%%%%%%%%%%%%%%%%%%%%%%%%%%%%%%%%%%%%%%%%%%%%%%%%%%%%%%%%%%%%%%%%%%%% 
(\ref{yf}). 
Introducing (\ref{yf}) into (\ref{cs}) we obtain

\ba      
f_-\d_-\d_-(f_-)=f_+\d_+\d_+(f_+)= A^2/2      
\ea     
where $A^2$ is a constant. Thus both $f_+$ and $f_-$ satisfy an identical     
equation:     
\ba     
ff'' =A^2/2 \la{ffa}     .
\ea
Note that $A=0$ implies $f''=0$ and thus we recover the condition (\ref{bgl}). 
We 
conclude that     
$A$ describes the deviation of the system from the uniform Hubble-like     
expansion.

Eq. (\ref{ffa}) can be solved multiplying by $f'$ and dividing by $f$:     
\ba     
[(f')^2]'= A^2 [\log f]'\;\rightarrow\;f'= A \sqrt{\log (f/H)} \la{ep}     
\ea     
where $H$ is an arbitrary constant.

Eq. (\ref{ep}) can be solved in the standard manner. We obtain     
\ba     
z-z_0=h\int_{F_0}^F \frac{dF'}{\sqrt{\log (F')}} \la{sol}     
\ea     
where     
we have introduced the notation     
\ba     
F=f/H\;;\;\;h=H/A
\ .
\ea

When (\ref{yf}) is introduced into (\ref{ep}) we obtain     
\eq     
g\d_+[\log p]&=&-\frac{ (1\!+\!g)    ^2}4\frac{F_+'}{F_+}+\frac{g^2\!-\!1   
}4\frac{F_-'}{F_+}
\nn     
g\d_-[\log p]&=&-\frac{ (1\!+\!g)    ^2}4\frac{F_-'}{F_-} +\frac{g^2\!-\!1   
}4\frac{F_+'}{F_-}
\ . \la{efp}     
\eqx     
From this we deduce     
\eq    
g\log p&=&-\frac{ (1\!+\!g)    ^2}4\log F_+     
+\frac{g^2\!-\!1   }4F_-'\int \frac{dz_+}{F_+}+ \Delta_-(z_-)\nn     
g\log p&=&-\frac{ (1\!+\!g)    ^2}4\log F_-     
+\frac{g^2\!-\!1   }4F_+'\int \frac{dz_-}{F_-}+ \Delta_+(z_-)\ . \la{eh}     
\eqx     
The integrals on the R.H.S. can be evaluated using (\ref{ep}). Indeed     
\ba     
\int \frac {dz}{F}=\int \frac {dF}{F} \frac1{F'}=h     
\int \frac {d\log F}{\sqrt{\log F}}=2h\sqrt{\log F}\ .     
\ea     
We thus obtain     
\ba     
g\log p=-\frac{ (1\!+\!g)    ^2}4\log (F_+F_-) +\frac{g^2\!-\!1   }2\sqrt{\log 
(F_+)
\log (F_-)}\ .      
\ea     
where, for the two equations (\ref{eh}) to be consistent with each other,     
 we had to take      
$\Delta_\pm(z_\pm)=-\frac{ (1\!+\!g)    ^2}4\log F_\pm$. This finally gives     
\ba     
p(z_+,z_-)=p_0\ \exp\left\{-\frac{ (1\!+\!g)    ^2}{4g}\left[l_+^2+l_-^2\right]
+\frac{g^2\!-\!1   }
{2g}     
\ l_+l_-   \right\} \la{pfin}     
\ea     
\ba
%%%%%%%%%%%%%%%%%%%%%%%%%%%%%%%%%%%%%%%%%%%%%%%%%%%%%%%%%%%%%%%%%%%%%%%%%%
\!\!\!\!\!\!\!\!\!\!\!\!\!\!\!\!\!\!\!\!\!\!\!\!\!\!\!\!\!\!\!\!\!\!\!
\!\!\!\!\!\!\!\!\!\!\!\!\!\!\!\!\!\!\!\!\!\!\!\!\!\!\!\!\!\!\!\!\!\!\!
\!\!\!\!\!\!\!\!\!\!\!y(z_+,z_-)=\frac12 \left(l_+^2-l_-^2\right)   \la{ylpm}   
%%%%%%%%%%%%%%%%%%%%%%%%%%%%%%%%%%%%%%%%%%%%%%%%%%%%%%%%%%%%%%%%%%%%%%%%%%
\ea
where, by definition      
\ba     
l_\pm(z_\pm) = \sqrt{\log(F_\pm)}\ . \la{lf}     
\ea  

\begin{figure}[t]
\centerline{\includegraphics[height=9cm]{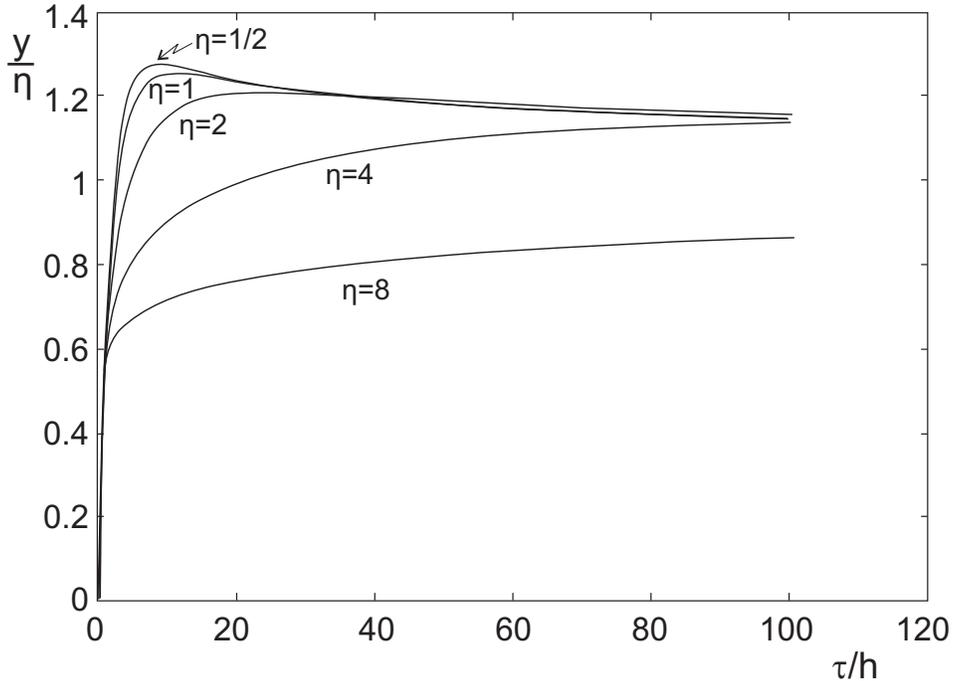}}
%%%%%%%%%%%%%%%%%%%%%%%%%%%%%%%%%%%%%%%%%%%%%%%%%%%%%%%%%%%%%%%%%%%%%
\caption{The ratio $y/\eta$ plotted versus proper-time $\tau/h$ for
 several values of space-time rapidity $\eta$. The boost-invariant picture  
corresponds to $y/\eta \equiv 1.$ The ``full-stopping'' initial condition is 
$y/\eta =0$ at $\tau=0.$ The asymptotic-value is still $1,$ but it is reached 
only at very large proper-times.}
%%%%%%%%%%%%%%%%%%%%%%%%%%%%%%%%%%%%%%%%%%%%%%%%%%%%%%%%%%%%%%%%%%%%%
\la{1}
\end{figure}

This completes the solution of hydrodynamic equations constrained by the     
generalized in-out cascade (\ref{yf}). One sees that the pressure     
depends on both $\tau$ and $\eta$. Thus the system is not     
boost-invariant but 
%%%%%%%%%%%%%%%%%%%%%%%%%%%%%%%%%%%%%%%%%%%%%%%%%%%%%%%%%%%%%%%%%%%%%%%%%%
 boost invariance is recovered in the limit     
$h\rightarrow 0$, $z_\pm$ fixed (see Eq. (\ref{sol})). On the other hand, 
it can 
be remarked that the Landau  asymptotic solution \cite{lan} can be recovered 
in 
the limit  $h$ fixed, $z_\pm \rightarrow \infty.$  
%%%%%%%%%%%%%%%%%%%%%%%%%%%%%%%%%%%%%%%%%%%%%%%%%%%%%%%%%%%%%%%%%%%%%%%%%%
%\footnote{Boost invariance is recovered in the limit     
%$h\rightarrow 0$, $z_\pm$ fixed (see Eq. (\ref{sol})).}.      
     
Other thermodynamic parameters are obtained from (\ref{esp}), giving     
\ba     
s\sim \epsilon^{g/ (1\!+\!g)    }=s_0\ 
\exp\left\{-\frac{1+g}{4}\left[l_+^2+l_-^2
\right]+     
\frac{g\!-\!1   }{2}\ l_+l_-   \right\}=s_0\ \exp{(-g\theta)}     
\la{s}     
\ea     
where we have denoted     
\ba     
\theta\equiv \log (T_0/T) =\frac{1+g}{4g}\left[l_+^2+l_-^2\right]-     
\frac{g\!-\!1   }{2g}\ l_+l_-   \la{theta}     
\ea     
with     
$s_0$ and $T_0$ denoting the entropy and temperature at the beginning of the     
hydrodynamic evolution.    

To illustrate the  deviation  of our solution (\ref{ylpm}) from the 
in-out Bjorken-Gottfried-Low Ansatz (\ref{bgl}), we show in the Fig.\ref{1} 
the ratio $y/\eta$ vs $\tau/h$ (the proper time measured in units of $h$) 
 for several (fixed) values of 
$\eta$. One sees that in the limit $\tau/h \rightarrow 0$ the rapidity $y$ 
vanishes. In this limit we have 
(\emph{e.g.} for the region $y\ge 0$)
\ba
y\approx (\tau/2h)^2 \sinh 2\eta =(t/2h)^2-(z/2h)^2\leq \frac14 (t/h)^2
\ .
\ea

Thus for a fixed (small) $t/h$ the fluid starts at rest  and acquires some 
 velocity
at later times, as in the Landau ``full stopping'' solution. At large times,
$\tau \rightarrow \infty$, one obtains $y \approx \eta$, i.e. the in-out Ansatz 
(\ref{bgl}) is approximately recovered. Thus our solution does indeed 
interpolate between the 
Landau and Bjorken hydrodynamics.

A last comment is in order. In all cases, the solution of the flow is also
 defined outside
the kinematical light-cone. Indeed, there is some flow of energy entering 
the light-cone from outside. It could be physically interpreted as mimicking 
energy sources on the light-cone (``leading particle effect''). However, the 
relevance of hydrodynamical models near the light-cone is questionable.

\section {Entropy density at freeze-out }     
     
The observable results of the model depend in an essential way on the     
assumed freeze-out surface. The point is that the densities $s$ and     
$\epsilon$ which enter the hydrodynamic equations are densities per unit     
volume {\it in the rest frame of the fluid}. But we are generally     
interested in the distribution of entropy $dS/dy$ and/or of energy     
$dE/dy$ densities per unit of rapidity, as these quantities are      
possible to measure. For given $s$ and $\epsilon$, $dS/dy$ and $dE/dy$     
depend on the hypersurface at which the hydrodynamic evolution stops and     
the fluid changes into particles (freeze-out surface). To fix attention,     
in the following we discuss the entropy density.

\subsection{ General freeze-out surface}
     
The evaluation of the entropy density per unit of rapidity for a given     
freeze-out surface can be performed in two steps.     
     
First we evaluate the amount of entropy in an infinitesimal volume      
along the freeze-out surface:     
\ba     
dS=s u^\mu d\sigma_\mu     
\ea     
where $u^\mu$ is the 4-velocity of the fluid and $d\sigma_\mu$     
is the 4-vector orthogonal to the surface satisfying     
\ba     
d\sigma^\mu d\sigma_\mu = dz^\mu dz_\mu =dz_+dz_-     
\ea     
Consider the (space-like) surface     
\ba     
\Phi(z_+,z_-)=C \la{phi}     
\ea     
where $C$ is a constant. Then      
\ba     
\Phi_+ dz_++\Phi_- dz_-=0\;;\;\; \Phi_\pm \equiv\frac{\d \Phi}{\d z_\pm}
\ .
\la{dif}     
\ea     
It follows that the unit vector orthogonal to the surface is     
\ba     
n_+= \frac{\Phi_-}{\sqrt{\Phi_+\Phi_-}}\;;\;      
n_-= \frac{\Phi_+}{\sqrt{\Phi_+\Phi_-}}\ .     
\ea     
The infinitesimal length along the surface is      
\ba     
d\sigma= \sqrt{dz_+dz_-}= dz_-\sqrt{\Phi_-/\Phi_+}=     
dz_+\sqrt{\Phi_+/\Phi_-}     
\ea     
     
Therefore      
\ba     
u^\mu d\sigma_\mu=[u^+\Phi_++u^-\Phi_-]\frac{dz_-}{2\Phi_+}=     
[u^+\Phi_++u^-\Phi_-]\frac{dz_+}{2\Phi_-} \ .\la{dsig}     
\ea     
     
In the second step we express the infinitesimal volume along the     
freeze-out surface in terms of the infinitesimal interval of rapidity.     
This can be done using the relation (\ref{sol}) which gives $z_\pm$     
as a function of $F_\pm=\exp{\left(l_\pm^2\right)}$. We have     
\ba     
dz_\pm = \frac {dz_\pm}{dF_\pm} \frac{dF_\pm}{dl_\pm^2} \     
dl_\pm^2=h\ \frac{\exp{\left(l_\pm^2\right)}} {l_\pm}\ dl_\pm^2     
\ea     
Using this relation and (\ref{sol}), the R.H.S. of (\ref{dsig})     
 can be expessed in terms of $l_+$ and $l_-$.

This in turn can be expressed in terms of rapidity $y$ using (\ref{yf})     
and the condition (\ref{phi}) which gives an additional relation      
between $z_+$ and $z_-$ and thus following (\ref{sol}) also between     
$l_+$ and $l_-$. In particular, using the differential forms,     
we have     
\ba     
dl_\pm^2= \frac{\pm 2\Phi_\mp\  e^{l_\mp^2}/l_\mp}     
{\Phi_+\exp{\left(l_+^2\right)}/l_+ \!+\!\Phi_-\exp{\left(l_-^2\right)}/l_-} 
dy\!\;\rightarrow\;\!     
dz_\pm= \frac{\pm 2h\Phi_\mp}     
{\Phi_+l_-e^{-l_-^2} \!+\!\Phi_-l_+e^{-l_+^2}} dy \la{dzdy}     
\ea     
and thus, finally     
\ba     
u^\mu d\sigma_\mu=\frac{u^+\Phi_++u^-\Phi_-}{\Phi_+l_-e^{-l_-^2}     
+\Phi_-l_+e^{-l_+^2}} dy =e^{\frac12(l_+^2+l_-^2)}     
\frac{\Phi_+e^y+\Phi_-e^{-y}}{l_-\Phi_+e^y+l_+\Phi_-e^{-y}}\ .\la{finus}     
\ea

\subsection {Freeze-out at a fixed time}     
      
\begin{figure}[t]
\centerline{\includegraphics[height=9cm]{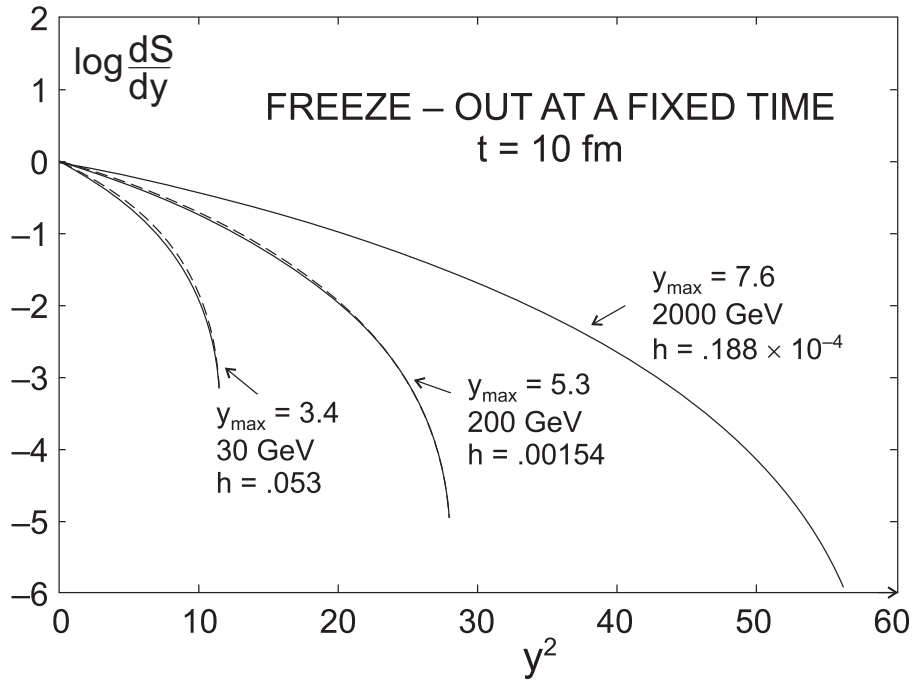}}
%%%%%%%%%%%%%%%%%%%%%%%%%%%%%%%%%%%%%%%%%%%%%%%%%%%%%%%%%%%%%%%%%%%%%%%%%%
 
\caption{The curve  $dS/dy$ with $t=const.$ (cf.  formula \eqref{dst}) compared     
with the Landau approximation (cf.  \eqref{slan}, dashed lines). The kinematical 
end-points at $y_{max}$ correspond to $z_-=0$ ($y\geq 0$).}
\la{2}
\end{figure}
%%%%%%%%%%%%%%%%%%%%%%%%%%%%%%%%%%%%%%%%%%%%%%%%%%%%%%%%%%%%%%%%%%%%%%%%%%

In this section we  take the surface  at      
 $t=const,$ for a first example. In the notation from the previous section we 
write     
\ba     
\Phi(z_+,z_-)= t = \frac12(z_++z_-)= const\;;\;\;\Phi_+=\Phi_-=\frac12     
\ea     
Using (\ref{finus}) we thus have     
\eq     
dS=s u^\mu d\sigma_\mu&=&     
\frac{ s\ e^{\frac12(l_+^2+l_-^2)}}{l_++l_-+(l_+-l_-)\tanh y}\ dy =\nn
&=&     
\frac{s_0\ e^{-(g\!-\!1   )(l_+-l_-)^2/4}}{l_++l_-+(l_+-l_-)\tanh y}\ dy
\ .\la{dst}     
\eqx     
     
If, following Landau, we approximate both $l_+$ and $l_-$ by large
constants, then for finite $y$ the difference $(l_+-l_-)$ is small and
we have 
     
\ba
dS \sim e^{-(g\!-\!1   )(l_+-l_-)^2/4} \la{slan} \ .    
\ea

 For  $g=3$  this formula is identical  to the  asymptotic result of Landau 
\cite{lan}. This can be seen when displaying the distribution $dS/dy;$  
In Fig.\ref{2}, one shows  $dS/dy$ with $t=const.$ (formula \eqref{dst})  
compared     
with the Landau approximation, formula (\ref{slan}), for different values of the 
parameter $h$, 
%%%%%%%%%%%%%%%%%%%%%%%%%%%%%%%%%%%%%%%%%%%%%%%%%%%%%%%%%%%%%%%%%%%%%%%%%%
which, by simple rescaling of the kinematic variables, correspond to different 
end-points in rapidity.
%%%%%%%%%%%%%%%%%%%%%%%%%%%%%%%%%%%%%%%%%%%%%%%%%%%%%%%%%%%%%%%%%%%%%%%%%%

One     
should keep in mind, however, that the relation $t=const.$ implies     
a different relation between $l_\pm$ and $y$ than the condition     
$\tau\sim const.$, which is the freeze-out condition considered\footnote{More 
precisely, Landau 
discusses the limitation of the 1+1 dimensional motion 
%%%%%%%%%%%%%%%%%%%%%%%%%%%%%%%%%%%%%%%%%%%%%%%%%%%%%%%%%%%%%%%%%%%%%%%%%%
by its transition to  the  3+1 dimensional hydrodynamical motion
%%%%%%%%%%%%%%%%%%%%%%%%%%%%%%%%%%%%%%%%%%%%%%%%%%%%%%%%%%%%%%%%%%%%%%%%%%
and shows that it appears at 
$\tau\sim const.$ playing the role of a freeze-out surface.} by Landau 
\cite{lan}. As discussed in the next subsection, this leads to a rather 
different  shape of the distribution $dS/dy$.

\subsection {Freeze-out at a fixed proper time}     
     
To investigate the relation to the Landau solution and its comparison 
with the Bjorken one we 
consider the freeze-out at a     
fixed proper time.      
     
Using the notation of Section 5.1 we have     
\ba     
\Phi(z_+,z_-)=z_+z_-=\tau^2=const.     
\ea     
\ba     
z_+dz_-+ z_-dz_+=0\;;\;\;\Phi_\pm=z_\mp     
\ea     
and thus     
\ba     
dS=he^{-(g\!-\!1   )(l_+-l_-)^2/4}     
\frac{e^yz_-+e^{-y}z_+}{l_-e^yz_-+l_+e^{-y}z_+}dy\ . \la{bjgen}     
\ea     
This is a general formula. When supplemented by (\ref{yf}) and      
(\ref{sol}), it expresses $dS$ in terms $y$ and $\tau$.     
     
When $h\rightarrow 0$ we can use the approximation (see the Appendix)      
\ba     
z_\pm=h \frac{F_\pm}{l_\pm} \la{appr1}     
\ea     
 to obtain     
\ba     
dS     
=h\ \frac{l_++l_-}{2l_+l_-}\ e^{-\f{g-1}{4}(l_+-l_-)^2}\ dy \ .\la{stauy}      
\ea     
For $l_\pm \rightarrow \infty$ and fixed $y$ one recovers the Landau result 
(\ref{slan}).

The  result given by (\ref{bjgen}) is plotted in Fig.\ref{3}  where $dS/dy$, is 
displayed for different values of the parameter h and compared with the 
approximate formula 
\ba
\frac{dS}{dy} =S_0\ e^{\sqrt{L^2-y^2}}  \la{sll}
\ea
 obtained by Landau \cite{lan}. The parameter $L$ was adjusted to obtain the 
correct slope at $y=0$.

 One observes some deviations from the perfect Gaussian which was considered  
in a simplified version \cite{carr} of the Landau model (and which agrees -if 
the 
multiplicity distribution is assumed  proportional to the entropy- with the data 
\cite{RH}).  Note that at fixed $\tau$ and  $h\to 0,$ the 
distribution becomes significantly flatter, going smoothly to the Bjorken limit 
at $h = 
0.$

\begin{figure}[t]
\centerline{\includegraphics[height=9cm]{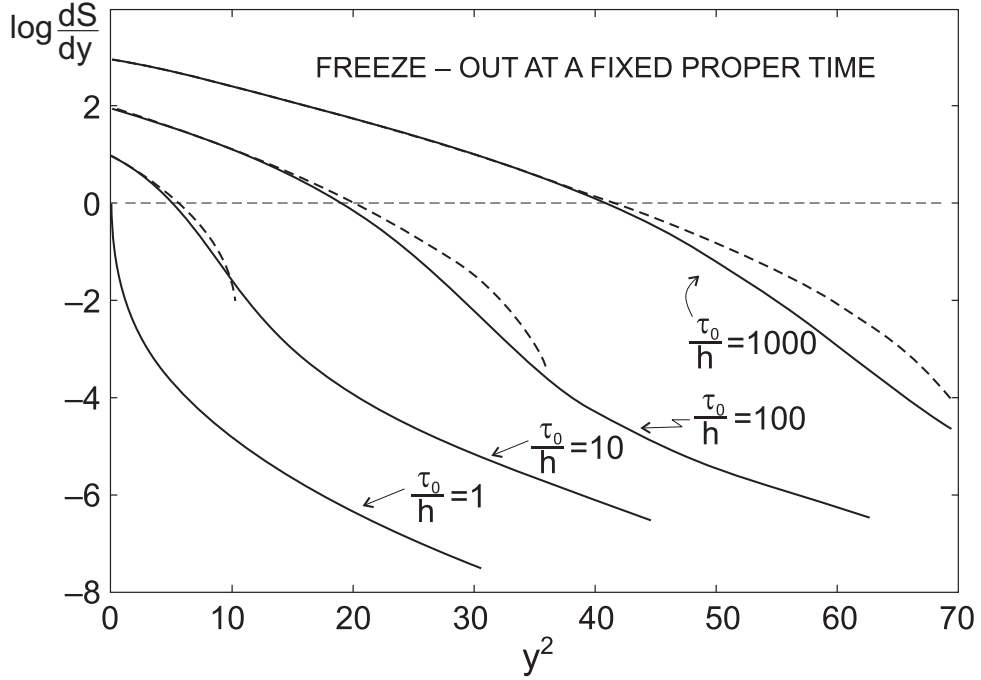}}
\caption{The curve  $dS/dy$ with $\tau=const.$  for various 
values of $\tau/h.$ Full lines: Eq.(\ref{stauy}); Dashed lines: Eq.(\ref{sll}). 
Note that, for clarity, the curves were shifted by $\log_{10}\tau/h$.}
\la{3}
\end{figure}

\subsection {Freeze-out at a fixed temperature}

 Instead of considering the freeze-out surface at the limit where the transverse 
motion becomes relevant (cf. \cite{lan}), a natural conjecture is to consider 
freeze-out at a fixed temperature, \emph{i.e.} when the temperature reaches the 
value 
where pions are expected to become liberated, \emph{e.g.} \cite{revhyd,also}.  
     
Denoting the initial temperature by $T_0$ and the freeze-out temperature     
 by $T_F$ we have     
\ba     
\Phi(z_+,z_-)\equiv \frac{1+g}{4}\left[l_+^2+l_-^2\right]-     
\frac{g\!-\!1   }{2}\ l_+l_-       
=g\log (T_0/T_F) \equiv g\theta = const. \la{phitheta}     
\ea     
Hence
\eq     
\Phi_+&=&\frac12[(g+1)l_+-(g\!-\!1   )l_-]l_+'=\frac{(g+1)l_+-(g\!-\!1   )l_-}     
{4h\exp{\left(l_+^2\right)}}\ \nn     
\Phi_-&=&\frac12[(g+1)l_--(g\!-\!1   )l_+]l_-'=\frac{(g+1)l_--(g\!-\!1   
)l_+}{4h\exp{\left(l_-^2\right)}}     
\eqx     
where we have used the relation following from (\ref{sol}):     
\ba     
l'=\frac{d(\sqrt{\log F})}{dz}=\frac{F'}{2F\sqrt{\log F}}
=\frac1{2hF}=
e^{-l^2}/2h
\ .
\ea     
Therefore using (\ref{s}) and (\ref{finus}) we have     
\ba     
dS\sim e^{-(g\!-\!1   ) (l_+-l_-)^2/4}\frac{l_++l_-}{(g+1)l_+l_-     
-\frac{g\!-\!1   }2(l_+^2+l_-^2)}
\ .
\ea

Now, the relations      
\ba     
\frac{1+g}{4}\left[l_+^2+l_-^2\right]-     
\frac{g\!-\!1   }{2}\ l_+l_-       
=g\theta ;;\;\; l_+^2-l_-^2 =2y \la{lty}     
\ea     
imply     
\ba     
l_-^2=\frac{g+1}2\theta-y+\frac{g\!-\!1   }2\sqrt{\theta^2-y^2/g}\;;\;\;     
l_+^2=2y+l_-^2     
\ea     
     giving     
\eq     
(g+1)l_+l_-     
-\frac{g\!-\!1   }2(l_+^2+l_-^2)&=&2g\sqrt{\theta^2-y^2/g}\;;\;\;\nn     
\frac{g\!-\!1   }4(l_+-l_-)^2&=&\frac{g\!-\!1   
}2\left[\theta-\sqrt{\theta^2-y^2/g}
\right]\nn     
l_++l_-&=&\sqrt{2}y\left[\theta-\sqrt{\theta^2-y^2/g}\right]^{-1/2}\ .     
\eqx   
Thus we finally  obtain      
\ba     
dS\sim e^{\frac{g\!-\!1   }2\left[\sqrt{\theta^2-y^2/g}-\theta\right]}     
\left(\theta-\sqrt{\theta^2-y^2/g}\right)^{-1/2}     
\frac{ydy}{\sqrt{\theta^2-y^2/g}}\ \la{temp}.     
\ea
 
One sees that this formula exibits a     
singularity\footnote{The singularity in $dS/dy$  does not come 
from a true singularity 
in the kinematical domain
but is due to a maximum value reached by the rapidity $y$ as a function of 
$z_-.$} by its transition to  the  3+1 dimensional hydrodynamical motion
%%%%%%%%%%%%%%%%%%%%%%%%%%%%%%%%%%%%%%%%%%%%%%%%%%%%%%%%%%%%%%%%%%%%%%%%%%
 at $y^2=g\theta^2$ which is of course unphysical and reflects the 
singular initial conditions of our approach and the expected limitation 
of hydrodynamic models to the more central scattering region.

\begin{figure}[htb]
\centerline{\includegraphics[height=9cm]{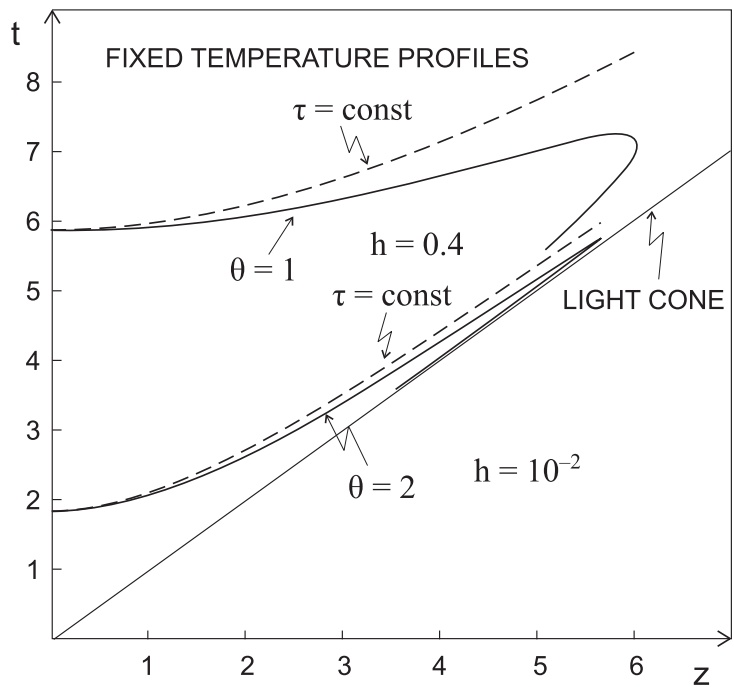}}
\caption{The profiles $\theta =\ const. $  for two values of 
$\theta= \log T_0/T$ (continuous lines). The comparison is made with fixed 
$\tau$ 
(dashed lines). Note that  a different value of the parameter h has been chosen 
to 
obtain a comparable  range in space-time.}
\la{4}\end{figure}  
It is, however, interesting to observe that the hypersurface $T=$const is only 
partly 
space-like. It becomes time-like at the rapidity      
 determined from the     
condition $\Phi_-=0$, i.e., $(g+1)l_-=(g\!-\!1   )l_+$, giving (c.f. 
(\ref{lty}))     
\ba     
y_m=\frac{2g}{g+1}\theta \;;\;\;\; l_+^2=(g+1)\theta .
\ea     
This is illustrated in Fig.\ref{4}, where two profiles $\theta=\ const.$ are 
displayed. 

\begin{figure}[t]
\centerline{\includegraphics[height=9cm]{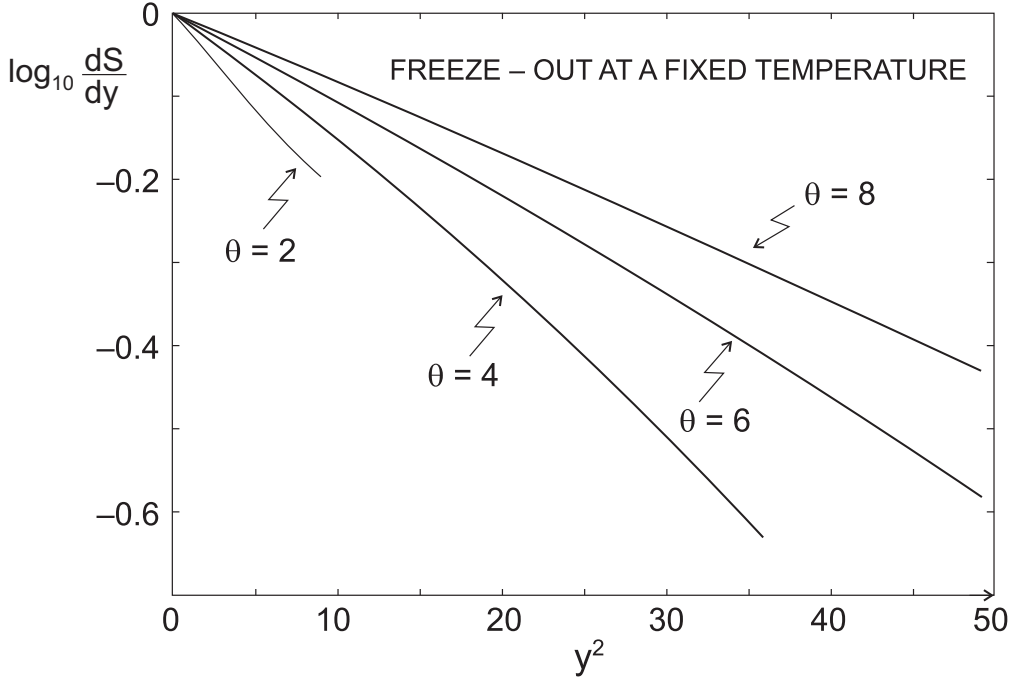}}
\caption{The curve  $dS/dy$ with $\theta=const.$  for various 
values of $\theta$, $y<3\theta/2$.}
\la{5}\end{figure}

The numerical estimates show that -for large enough $\theta$ ($\theta     
\geq 2$)- the effect of the singularity on the entropy distribution as 
a function of $y$ is not 
significant in the  region $y\leq y_m$.  This is shown in Fig.\ref{5} where 
$dS/dy$ is     
plotted for several values of $\theta \geq 2$. One sees that all     
distributions are close to Gaussians. 
 The slope, however, is rather     
small, certainly smaller than required by data, unless one considers a larger 
value of 
the parameter $g,$ \emph {i.e.} a smaller value of the speed of sound.

%It is useful to compare the isothermal freeze-out surface with the 
%corresponding  
%isochronic (in proper-time) surface corresponding to the coupling of tranverse 
%motion 
%\cite{lan}. We notice that when $\theta$ increases, the two curves  are very 
%close 
%until the temperature surface reaches the turning point when its gradient 
%becomes 
%time-like, which is just the limitation we have considered. At smaller $\theta$ 
%the 
%%difference is larger, as already noticed in the numerical simulations of 
%\cite{also}. 
     
\section{ Conclusions and outlook}     
     
We have investigated longitudinal hydrodynamic expansion of a perfect     
fluid forming an     
infinitely thin layer at the initial time and satisfying the equation of state 
with an arbitrary sound velocity.      We proposed a generalized in-out Ansatz 
which unifies the Landau and Bjorken approaches to relativistic
hydrodynamics.
     
Our conclusions can be summarized as follows.     
     
(i) When the Bjorken-Gottfried-Low in-out hypothesis (\ref{bgl}) is     
assumed, the hydrodynamics implies that all thermodynamic properties of     
the fluid depend only on proper time, i.e. the system is     
boost-invariant.     
     
(ii) We proposed a generalized in-out hypothesis, (\ref{yf}), and     
discussed its implications. It turns out that in this case the     
hydrodynamic equations can be solved in an analytic form, giving     
explicit formulae for the thermodynamic characteristics of the fluid in     
terms of their initial values and one free parameter $h$ defining a     
dynamical scale in  configuration space.      
     
(iii) The resulting entropy distribution in rapidity, $dS/dy$, was     
evaluated and shown to depend significantly on the assumed condition for 
the 
freeze-out.

(a) For freeze-out at a fixed proper-time the density is close to
%%%%%%%%%%%%%%%%%%%%%%%%%%%%%%%%%%%%%%%%%%%%%%%%%%%%%%%%%%%%%%%%%%%%%%%%%% 
(but with some deviation, particularly at small $\tau,$ from)
%%%%%%%%%%%%%%%%%%%%%%%%%%%%%%%%%%%%%%%%%%%%%%%%%%%%%%%%%%%%%%%%%%%%%%%%%% 
  a Gaussian which is traditionally attributed to Landau solution.      
It  tends smoothly to the Bjorken boost-invariant     
solution for $h\rightarrow 0$.     
     
(b) For freeze-out at a fixed temperature the distribution is even
 closer to a    
Gaussian, if one restricts to the region where the freeze-out surface is space 
like and if  the ratio $T_0/T_F$ is large enough.    

(iv) It is worthwhile to note that the freedom in the choice of the
value of the 
sound velicity may be helping in phenomenological applications of these results 
to data.  
     
Compared to other recent $(1\!+\!1)$ hydrodynamical 
models \cite{csorgo, pratt}, our solution 
is mainly characterized  by the smooth, one-parameter dependent transition
 between
 the Bjorken and Landau hydrodynamical models and by its analytic simplicity.
  It would be useful 
 to study further the classification of all the solutions in a unified
 framework.
 
 More generally,      
there is a clear need for an extension of our investigation to include more 
flexible initial conditions,
 relaxing the point-like character of the fluid at the beginning of the 
evolution.
This, however, demands a more     
sophisticated analysis (e.g. an application of the general recipe of 
\cite{khal}) and goes beyond the scope of this paper. 

%%%%%%%%%%%%%%%%%%%%%%%%%%%%%%%%%%%%%%%%%%%%%%%%%%%%%%%%%%%%%%%%%%%%%%%%%%%%%%%%
On a theoretical ground, it would be interesting to have a physical 
interpretation of the generalized Ansatz \eqref{yf}, which appears as a 
mathematical harmonic property $\d_+\d_-y=0$ of the hydrodynamical flow. In 
particular, an extension to this case of the Gauge/Gravity correspondence 
applied in Ref.\cite{janik} to the Bjorken Boost-invariant flow, would be 
insightful. 
%%%%%%%%%%%%%%%%%%%%%%%%%%%%%%%%%%%%%%%%%%%%%%%%%%%%%%%%%%%%%%%%%%%%%%%%%%%%%%%%
     
 \eject    
     
\section*{Appendix. Solution of the equation (\ref{sol})}     
     
We rewrite (\ref{sol}) as     
\ba     
(z-\zeta)= h\int_{F_0}^F \frac {dv}{\sqrt{\log v}}      
\ea     
where $F= f/H$ and $h=H/A$.     
     
Changing the variable of integration:     
\ba     
\log v =u^2\;;\;\; 2udu =dv/v\;;\;\; dv/u=2 e^{u^2}du     
\ea     
we arrive at 
\ba     
z-\zeta= 2hF     
\int^{\sqrt{\log F}}_{\sqrt{\log F_0}} e^{u^2-\log F} du =     
2\frac{HF}{A}\left[D\left(\sqrt{\log F}\right)-     
D\left(\sqrt{\log F_0}\right)\right]\la{zet}     
\ea     
This integral is the so called Dawson's integral: $D(x)= e^{-x^2} \int_0^x     
e^{u^2}du$. For large $\sqrt{\log F}$     
it approaches $(2\sqrt{\log F})^{-1}$ and thus we obtain      
\ba      
z-\zeta\approx \frac{HF}{A\sqrt{\log F}} \la{zfp}     
\ea     
The asymptotic expansion of $D(x)$ is      
\ba     
D(x)= \frac1{2x}\sum_{n=0}^\infty \frac{\Gamma(n+1/2)}{\Gamma(1/2)}\frac1     
{x^{2n}}     
\ea     
     
For small $x$ one can evaluate this integral effectively by     
the series     
expansion:     
\ba     
D(x)=e^{-x^2}\int_0^x \sum_{n=0}^\infty \frac {x^2n}{n!}=     
x e^{-x^2} \sum_{n=0}^\infty \frac {x^{2n}}{(2n+1)n!} \la{dx}     
\ea     
Thus we have in this case     
\ba     
z-z_0=2\frac{H}{A} \left[\sqrt{\log (F)}      
\sum_{n=0}^\infty \frac {(\log F)^n}{(2n+1)n!}-     
\sqrt{\log (F_0)}      
\sum_{n=0}^\infty \frac {(\log F_0)^n}{(2n+1)n!}\right]     
\ea     
\eject     
     
%\vspace{0.3cm}     
\subsection*{Acknowledgements}     
     
We thank Wojtek Florkowski, Jean-Yves Ollitrault and Kacper Zalewski for useful 
comments.     
This investigation was partly supported by the     
MEiN research grant 1 P03B 045 29 (2005-2008) and by 6 Program of European
Union ``Marie Curie Transfer of Knowledge'' Project: Correlations in Complex
Systems ``COCOS'' MTKD-CT-2004-517186.

\end{document}